\documentclass[article]{Definitions/mdpi}
\usepackage{mathrsfs}
\usepackage{relsize}
\usepackage[utf8]{inputenc}
\firstpage{1} 
\pubvolume{1}
\issuenum{1}
\articlenumber{0}
\pubyear{2025}
\copyrightyear{2025}
\datereceived{ } 
\daterevised{ } % Comment out if no revised date
\dateaccepted{ } 
\datepublished{ } 
\hreflink{https://doi.org/} % If needed use \linebreak
\Title{Turbo Coded Single Sideband OFDM-OQAM Signaling through Frequency
       Selective Rayleigh Fading Channels}

\TitleCitation{Title}

\Author{Kasturi Vasudevan $^{1}$\orcidA{}}

\AuthorNames{Kasturi Vasudevan}

\isAPAStyle{%
       \AuthorCitation{Vasudevan, K.}
         }{%
        \isChicagoStyle{%
        \AuthorCitation{Lastname, Firstname, Firstname Lastname, and Firstname Lastname.}
        }{
        \AuthorCitation{Vasudevan, K.}
        }
}

\address{%
$^{1}$ \quad IIT Kanpur; vasu@iitk.ac.in\\
}

\corres{Correspondence: vasu@iitk.ac.in;
        }

\abstract{
This work investigates the bit-error-rate (BER) performance of turbo coded
orthogonal frequency division multiplexed - offset quadrature amplitude modulated (OFDM-
OQAM) signals transmitted through frequency selective Rayleigh fading channels in the 
presence of carrier frequency offset (CFO) and additive white Gaussian noise (AWGN). The 
highlight of this work is to use the root raised cosine (RRC) pulse and its Hilbert 
transform as the complex-valued transmit filter and a simple matched filter at the 
receiver. The proposed system is similar to single sideband (SSB) modulation, that
has roots in analog communications. Turbo code and subcarrier diversity is
employed to improve the BER performance over that of an uncoded system.
Discrete-time algorithms for frame detection, two-step CFO,
channel and noise variance estimation have been proposed. A
single transmit and receive antenna is assumed. Similar work has not been
done earlier.
}


\begin{document}
\nolinenumbers
%%%%%%%%%%%%%%%%%%%%%%%%%%%%%%%%%%%%%%%%%%
%\setcounter{section}{-1} %% Remove this when starting to work on the template.
%\section{How to Use this Template}

%The template details the sections that can be used in a manuscript. Note that the order and names of article sections may differ from the requirements of the journal (e.g., the positioning of the Materials and Methods section). Please check the instructions on the authors' page of the journal to verify the correct order and names. For any questions, please contact the editorial office of the journal or support@mdpi.com. For LaTeX-related questions please contact latex@mdpi.com.%\endnote{This is an endnote.} % To use endnotes, please un-comment \printendnotes below (before References). Only journal Laws uses \footnote.

% The order of the section titles is different for some journals. Please refer to the "Instructions for Authors” on the journal homepage.

\section{Introduction}

%The introduction should briefly place the study in a broad context and highlight why it is important. It should define the purpose of the work and its significance. The current state of the research field should be reviewed carefully and key publications cited. Please highlight controversial and diverging hypotheses when necessary. Finally, briefly mention the main aim of the work and highlight the principal conclusions. As far as possible, please keep the introduction comprehensible to scientists outside your particular field of research. Citing a journal paper \citep{ref-journal}.  Now citing a book reference \citep{ref-book1,ref-book2} or other reference types \citep{ref-unpublish,ref-url}. Please use the command \citep{ref-proceeding,ref-thesis} for the following MDPI journals, which use author--date citation: Administrative Sciences, Arts, Behavioral Sciences, Businesses, Econometrics, Economies, Education Sciences, European Journal of Investigation in Health, Psychology and Education, Games, Genealogy, Histories, Humanities, Humans, IJFS, Journal of Intelligence, Journalism and Media, JRFM, Languages, Laws, Literature, Psychology International, Publications, Religions, Risks, Social Sciences, Tourism and Hospitality, Youth. 

Orthogonal frequency division multiplexing-offset
quadrature amplitude modulation (OFDM-OQAM) is a modulation
of choice for 5G and beyond technologies due to its spectral
containment and robustness to carrier frequency offset (CFO), compared to
classical OFDM. However, it has been shown in
\cite{6663392,Vasudevan2015,Vasu_Adv_Tele_2017} that CFO in OFDM can be
estimated very accurately with linear complexity and large scope for parallel
processing. Moreover, the spectrum of OFDM can also be contained using
pulse shaping \cite{Vasu_Book10}. The overhead due to the cyclic prefix (CP) in
OFDM can be reduced by increasing the ``frame'' size. In fact it is the
``training'' or ``pilot'' that increases the overhead, rather than CP
\cite{6663392,Vasudevan2015,Vasu_Adv_Tele_2017}.
Most of the literature on OFDM-OQAM or its variants like filter bank
multicarrier (FBMC) or universal filtered multicarrier (UFMC), use
non-Nyquist pulses like isotropic orthogonal transfer algorithm (IOTA)
and extended Gaussian functions (EGF), which result in intersymbol
interference (ISI)
\cite{5753092,6852083,7509396,7564682,8014377,9339835}.
A brief survey of OFDM-OQAM is given below.

Channel estimation for coherent optical FBMC-OQAM is given in \cite{9985452}.
Underwater acoustic communications using FBMC-OQAM is presented in \cite{9993782}.
FBMC for massive multiple input multiple output (MIMO)
systems is investigated in \cite{10088421,10439008}.
Blind joint CFO and sampling time offset (STO) estimation for FBMC-OQAM
systems is addressed in \cite{10176180}.
Downlink precoding for cell-free FBMC/OQAM systems with asynchronous
reception is considered in \cite{10178073}.
FBMC-OQAM transceivers for 6G and beyond has been reviewed in \cite{10210617}.
A low-complexity symbol reconstruction based on direct symbol decision
for FBMC-OQAM systems is presented in \cite{10304162}.
The performance of FBMC and OFDM in fiber optic systems is studied in
\cite{10354509}.
A nonuniform FBMC-OQAM with subchannels having arbitrary bandwidths is
discussed in \cite{10438945}. Voice transmission over
voiceband channels using FBMC-OQAM is presented in \cite{10586986}. Design of
waveforms for FBMC-OQAM is considered in \cite{10910057,11014766}.
Peak-to-average power ratio (PAPR) reduction in FBMC-OQAM is given in
\cite{10924171}. Detection of OFDM-OQAM signals in the presence of phase
noise is discussed in \cite{11087565}.
A new pilot pattern for channel estimation in FBMC systems is considered in
\cite{11122462}.
 
In \cite{9187991}, an efficient channel estimation for UFMC is proposed.
In \cite{9211500}, it is shown that the BER performance of full duplex (FD)
FBMC (FD-FBMC) is better than FD-UFMC and FD-OFDM.
A reliability-based signal detection for UFMC is proposed in \cite{9290058}.
Sparse code multiple access in UFMC is presented in \cite{9519162}.
A performance comparison between OFDM and UFMC for passive optical networks
(PON) is studied in \cite{9541102}.
A deep learning-aided signal detection for two-stage index modulated
UFMC systems is proposed in \cite{9502166}.
Multiple access techniques for OFDM, FBMC and UFMC is presented in
\cite{9713895}.
Blind carrier frequency offset (CFO) estimation in UFMC is investigated in
\cite{9695339,9718522}.
Performance comparison of FBMC, UFMC and filtered OFDM (F-OFDM) is given
in \cite{9861227}.
UFMC for MIMO systems is addressed in \cite{10583879}.
Peak-to-average power ratio (PAPR) reduction in OFDM,
F-OFDM and UFMC is discussed in \cite{10870287}.
UFMC over optical fiber networks for 6G is investigated in \cite{11192529}.
A low complexity UFMC receiver for 6G is given in \cite{11195078}.

In this work, we use the
pulse having the root raised cosine (RRC) spectrum and its modified
Hilbert transform as the complex-valued transmit filter. This approach also
corresponds to single sideband (SSB) modulation that has roots in analog
communications. Therefore, the bandwidth of the transmitted signal is half that
of conventional QAM or OQAM systems and is narrow enough such that the channel
can be considered to be distortionless, justifying the use of a matched filter
receiver \cite{Vasu_Book10,KV_SIPS2023,1d40624a-9f4c-4721-aafb-a4aeee8e807f}.
In practice however, the distortionless channel assumption may not strictly
hold, which results in ISI.
The bit-error-rate (BER) performance of
the proposed system has been analyzed using
computer simulations in the presence of carrier frequency offset (CFO),
frequency selective Rayleigh fading channel and additive
white Gaussian noise (AWGN). Turbo coding and subcarrier diversity has been
introduced to significantly improve the BER performance over that of the uncoded
system. Discrete-time algorithms for frame detection, CFO, channel
and noise variance estimation have been
discussed. Similar work has not been done earlier.

This paper is organized as follows. Section~\ref{Sec:System_Model} presents
the system model. In section~\ref{Sec:Rx_Algo}, we describe the discrete-time
receiver algorithms. Section~\ref{Sec:Results} contains the simulation results.
Finally, the conclusions are in section~\ref{Sec:Conclude}.

The notation used is as follows. Complex quantities are denoted by tilde,
e.g. $\tilde{x}$. Caret is used to denote estimate \cite{Vasu_Book10}
or Hilbert transform \cite{Haykin83}, depending on the context,
e.g. $\hat{x}$ denotes estimate or Hilbert transform of $x$. However,
we emphasize that by default, caret denotes an estimate. If it denotes
Hilbert transform, then it is explicitly stated. Boldface is used
for vectors and matrices. The letter ``$\mathrm{j}$'' denotes $\sqrt{-1}$,
``$\star$'' denotes convolution and ``$E[\cdot]$'' is the expectation
operator. ``$\Re\{\cdot\}$'' denotes the real part and ``$\Im\{\cdot\}$''
denotes imaginary part of a complex quantity.
%%%%%%%%%%%%%%%%%%%%%%%%%%%%%%%%%%%%%%%%%%%%%%%%%%%%%%%%%%%%%%%%%%%%%%%%%%%%%%%%
\section{System Model}
\label{Sec:System_Model}
%*******************************************************************************
\begin{figure}[tbhp]
\begin{center}
%\centering
\input{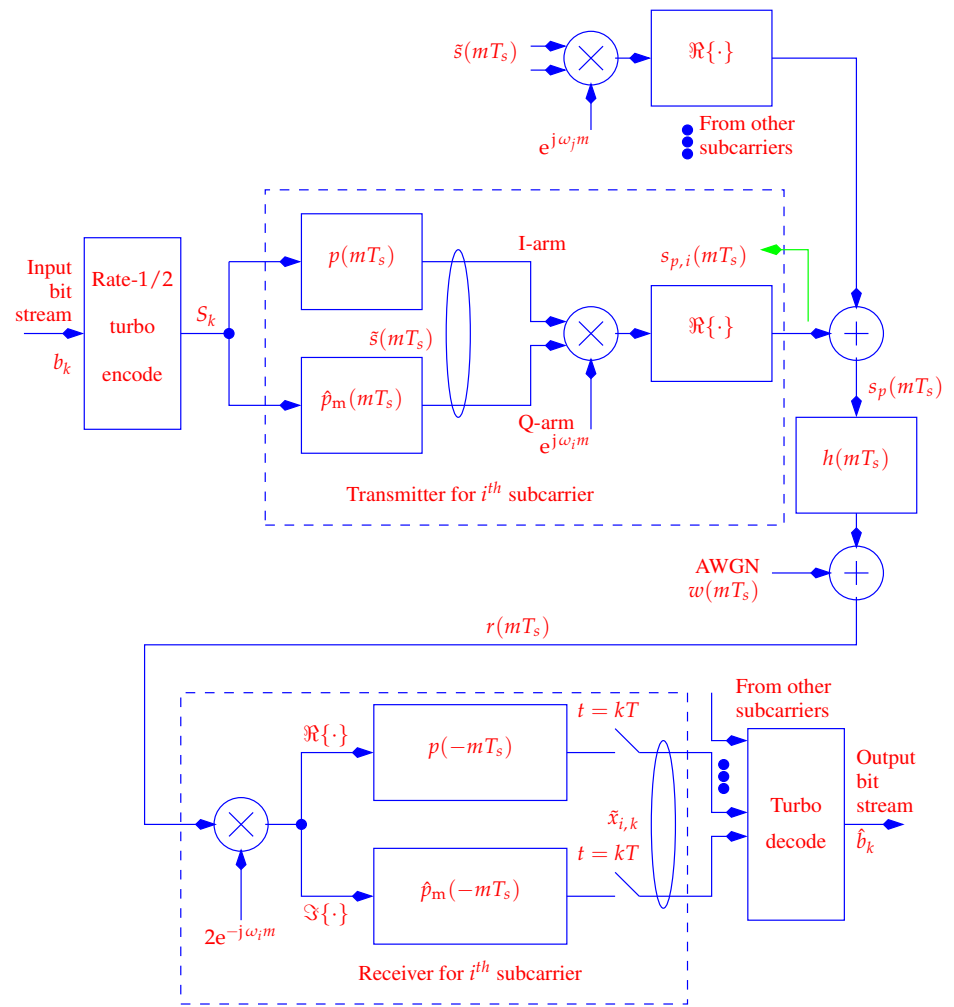}
\caption{Discrete-time system model of OFDM-OQAM.}
\label{Fig:OFDM_OQAM_Sys_Model1}
\end{center}
\end{figure}
\begin{figure}[tbhp]
\begin{center}
%\centering
\input{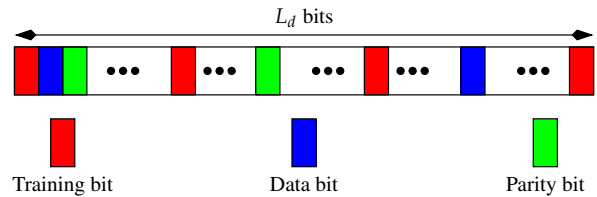}
\caption{Proposed frame structure of OFDM-OQAM.}
\label{Fig:OFDM_OQAM_Frame1}
\end{center}
\end{figure}

%*******************************************************************************
The block diagram of the proposed discrete-time OFDM-OQAM system is shown in
Figure~\ref{Fig:OFDM_OQAM_Sys_Model1}. Here subcarrier diversity is used,
that is, the same information is transmitted over multiple subcarriers
to improve the BER performance. A single transmit and receive antenna
(single input single output (SISO)) is assumed. The complex baseband
signal at the transmitter is given by \cite{KV_SIPS2023}
%*******************************************************************************
\begin{equation}
\label{Eq:Pap19_Eq1}
\tilde{s}(mT_s) = \sum_{k=0}^{L_d-1}
                   S_k
                  \tilde{p}_{\mathrm{m}}(mT_s - kT)
\end{equation}
%*******************************************************************************
where $L_d$ is the frame length in bits, $1/T$ is the symbol rate, $1/T_s$
is the sampling frequency and $\tilde{p}_{\mathrm{m}}(\cdot)$ is the
complex-valued transmit filter given by \cite{KV_SIPS2023}
%*******************************************************************************
\begin{equation}
\label{Eq:Pap19_Eq2}
\tilde{p}_{\mathrm{m}}(mT_s)=p(mT_s)+\mathrm{j}\,\hat{p}_{\mathrm{m}}(mT_s)
\end{equation}
%*******************************************************************************
where $p(\cdot)$ is the pulse having the root-raised cosine (RRC) spectrum and
$\hat{p}_{\mathrm{m}}(\cdot)$ is the pulse corresponding to the modified
Hilbert transform of the RRC (mHT-RRC) spectrum \cite{KV_SIPS2023}. We assume
that the interpolation factor is \cite{KV_SIPS2023}
%*******************************************************************************
\begin{equation}
\label{Eq:Pap19_Eq3}
\frac{T}{T_s} = I \quad \mbox{(an integer)}.
\end{equation}
%*******************************************************************************
The spectrum of the signal in (\ref{Eq:Pap19_Eq1}) extends over
\cite{Vasu_Book10}
%*******************************************************************************
\begin{align}
\label{Eq:Pap19_Eq3_1}
 0                & \le \omega \le \frac{\pi(1+\rho)}{I}   \nonumber  \\
 \Rightarrow
 \mbox{bandwidth} & = \frac{\pi(1+\rho)}{I}
\end{align}
%*******************************************************************************
where $\omega$ is the digital frequency in radians and $0<\rho \le 1$ is the
roll-off factor of the RRC spectrum. Thus (\ref{Eq:Pap19_Eq1}) is a single
sideband (SSB) signal.
Since a rate-$1/2$ turbo code is used, the frame contains both data and
parity bits, which are mapped to BPSK with amplitude $S_k\in \pm 1$.
We assume that bit 0 maps to $+1$ and bit 1 maps to $-1$. The frame
also contains training bits, which are distributed at random time slots
over the frame and spans the entire frame. Therefore, the first and last bits
of the frame correspond to training. This is illustrated in
Figure~\ref{Fig:OFDM_OQAM_Frame1}.
%*******************************************************************************
%*******************************************************************************
The number of data bits in a frame is
%*******************************************************************************
\begin{equation}
\label{Eq:Pap19_Eq4}
L_{d1}=L_d/3
\end{equation}
%*******************************************************************************
since the frame contains equal number of data, parity and training bits. The
generating matrix of the constituent encoder of the turbo code is given by
\cite{Vasu_Book10,6663392,Vasudevan2015,Vasu_ICWMC2016,
Vasu_Adv_Tele_2017,KV_OpSigPJ2019,Vasu_intech:2019,KV_ARCI2021,
KV_Oct_2021,Vasudevan23,KV_SIPS2023}
%*******************************************************************************
\begin{equation}
\label{Eq:Pap19_Eq4_1}
\mathbf{G}(D) = \left[
                \begin{array}{cc}
                 1    &  \frac{1+D^2}{1+D+D^2}
                \end{array}
                \right].
\end{equation}
%*******************************************************************************
The passband signal for the $i^{th}$ subcarrier is given by
%*******************************************************************************
\begin{equation}
\label{Eq:Pap19_Eq5}
 s_{p,\, i}(mT_s) = \Re
                    \left\{
                    \tilde{s}(mT_s)
                    \mathrm{e}^{\mathrm{j}\, \omega_i m}
                    \right\}
\end{equation}
%*******************************************************************************
where $\omega_i$ is the digital carrier frequency given by
%*******************************************************************************
\begin{equation}
\label{Eq:Pap19_Eq6}
\omega_i = \frac{2\pi k_i}{I} + \delta_i 
\end{equation}
%*******************************************************************************
where $1\le k_i < I/2$ is an integer and $\delta_i$ is the frequency offset
in the $i^{th}$ subcarrier, in radians. Note that $k_i=0$ corresponds to
the baseband signal (no modulation), and is hence not used.
In this work, we assume that the
frequency offset is at most $\pm 1\%$ of the symbol rate ($1/T$).
This implies that
%*******************************************************************************
\begin{equation}
\label{Eq:Pap19_Eq7}
\delta_i \in \left[\frac{-\pi}{50I},\, \frac{\pi}{50I}
             \right]
\end{equation}
%*******************************************************************************
where $I$ is given in (\ref{Eq:Pap19_Eq3}). Let
%*******************************************************************************
\begin{equation}
\label{Eq:Pap19_Eq7_1}
\delta_{\mathrm{max}} = \frac{\pi}{50I}.
\end{equation}
%*******************************************************************************
We assume that the frequency offset $\delta_i$ is uniformly distributed in
$[-\delta_{\mathrm{max}},\, \delta_{\mathrm{max}}]$.
The overall discrete-time
transmitted passband signal is given by
%*******************************************************************************
\begin{equation}
\label{Eq:Pap19_Eq8}
 s_p(mT_s) = \sum_{i=1}^{N_{\mathrm{sc}}}
              s_{p,\, i}(mT_s)
\end{equation}
%*******************************************************************************
where $N_{\mathrm{sc}}$ denotes the number of subcarriers
(subcarrier diversity). The discrete-time
signal at the receiver input is
%*******************************************************************************
\begin{align}
\label{Eq:Pap19_Eq9}
 r(mT_s) & =  s_p(mT_s) \star h(mT_s) + w(mT_s)    \nonumber  \\
         & = \sum_{l=0}^{L_h-1}
              h(lT_s) s_p(mT_s-lT_s) + w(mT_s)
\end{align}
%*******************************************************************************
where $h(\cdot)$ denotes the impulse response of the real-valued, quasi-static,
frequency selective fading channel, $L_h$ denotes the channel length and
$w(mT_s)$ denotes samples of real-valued AWGN with zero-mean and variance
$\sigma^2_w$. The ``quasi-static'' assumption implies that the channel
impulse response is time-invariant over one frame and varies independently
from frame-to-frame. The channel coefficients are real-valued,
wide sense stationary (WSS),
independent, zero-mean Gaussian random variables with variance $1/L_h$, that is
%*******************************************************************************
\begin{align}
\label{Eq:Pap19_Eq9_1}
 E
\left[
 h(lT_s)h(lT_s-mT_s)
\right] & = \frac{\delta_K(mT_s)}{L_h}              \nonumber  \\
        &   \stackrel{\Delta}{=}  R_{hh}(mT_s)
\end{align}
%*******************************************************************************
where $\delta_K(\cdot)$ is the Kronecker-delta function
\cite{Proakis92,Vasu_Book10}.
Assuming a flat fading channel for each subcarrier, which is true if the
signal bandwidth around the subcarrier is sufficiently narrow ($I$ in
(\ref{Eq:Pap19_Eq3_1}) is sufficiently large) \cite{Vasu_Book10,KV_SIPS2023},
the signal at the matched filter output for the $i^{th}$ subcarrier at the
$k^{th}$ symbol instant can be written as
%*******************************************************************************
\begin{equation}
\label{Eq:Pap19_Eq10}
\tilde{x}_{i,\, k} = \tilde{H}_i S_k (1+\mathrm{j}) + \tilde{z}_{i,\, k}
\end{equation}
%*******************************************************************************
where $\tilde{z}_{i,\, k}$ denotes samples of complex-valued AWGN and
%*******************************************************************************
\begin{align}
\label{Eq:Pap19_Eq11}
\tilde{H}(\omega) & = \sum_{l=0}^{L_h-1}
                       h(lT_s)
                      \mathrm{e}^{-\mathrm{j}\,\omega l}   \nonumber  \\
\Rightarrow
\tilde{H}_i       &   \stackrel{\Delta}{=}
                      \tilde{H}(\omega_i) 
\end{align}
%*******************************************************************************
is the discrete-time Fourier transform (DTFT) of the channel
\cite{Proakis92,Vasu_Book10} and $\omega_i$ is given by (\ref{Eq:Pap19_Eq6}).
Since $\delta_i$ in (\ref{Eq:Pap19_Eq7}) is small, $\omega_i$ in
(\ref{Eq:Pap19_Eq6}) can be approximated as
%*******************************************************************************
\begin{equation}
\label{Eq:Pap19_Eq11_1}
\omega_i \approx \frac{2\pi k_i}{I}.
\end{equation}
%*******************************************************************************
Using (\ref{Eq:Pap19_Eq9_1}) and (\ref{Eq:Pap19_Eq11_1}) we get
%*******************************************************************************
\begin{align}
\label{Eq:Pap19_Eq11_2}
 E
\left[
\tilde{H}_i
\tilde{H}_j^*
\right] & = \frac{1}{L_h}
            \sum_{m=0}^{L_h-1}
            \mathrm{e}^{-\mathrm{j}\, 2\pi m (k_i-k_j)/I}  \nonumber  \\
        & = \delta_K(k_i-k_j)
\end{align}
%*******************************************************************************
provided
%*******************************************************************************
\begin{equation}
\label{Eq:Pap19_Eq11_3}
k_i - k_j = \frac{In}{L_h}
\end{equation}
%*******************************************************************************
where $n$ is an integer.
Thus (\ref{Eq:Pap19_Eq11_3}) specifies the subcarrier spacing required for
uncorrelated fading, which is necessary for achieving the minimum BER.
Assuming that the filters $p(mT_s)$ and $\hat{p}_{\mathrm{m}}(mT_s)$
have unit energy, the one-dimensional variance of $\tilde{z}_{i,\, k}$ in
(\ref{Eq:Pap19_Eq10}) is equal to \cite{Vasu_Book10}
%*******************************************************************************
\begin{equation}
\label{Eq:Pap19_Eq12}
\frac{1}{2}
 E
\left[
\left|
\tilde{z}_{i,\, k}
\right|^2
\right] = 2\sigma^2_w.
\end{equation}
%*******************************************************************************
Since the information contained in (\ref{Eq:Pap19_Eq10}) is
$1/(2N_{\mathrm{sc}})$ bits, the average SNR per bit is given by
\cite{KV_OpSigPJ2019}
%*******************************************************************************
\begin{align}
\label{Eq:Pap19_Eq13}
\mathrm{SNR}_{\mathrm{av},\, b} & = \frac{2
                                         \times 2 N_{\mathrm{sc}}
                                         \times
                                          E
                                         \left[
                                         \left|
                                         \tilde{H}_i
                                         \right|^2
                                         \right]
                                          E
                                         \left[
                                          S_k^2
                                         \right]}
                                         {E
                                         \left[
                                         \left|
                                         \tilde{z}_{i,\, k}
                                         \right|^2
                                         \right]
                                         }         \nonumber  \\
                                & = \frac{N_{\mathrm{sc}}}{\sigma^2_w}
\end{align}
%*******************************************************************************
where we have used (\ref{Eq:Pap19_Eq11_2}) and (\ref{Eq:Pap19_Eq12}).
In the next section, we discuss the receiver algorithms.
%%%%%%%%%%%%%%%%%%%%%%%%%%%%%%%%%%%%%%%%%%%%%%%%%%%%%%%%%%%%%%%%%%%%%%%%%%%%%%%%
\section{Receiver Algorithms}
\label{Sec:Rx_Algo}
\subsection{Frequency Offset and Frame Detection}
\label{SSec:FOFF_Fr_Detect}
The first step is to detect the presence of a frame. Assuming a worst case
frequency offset of $\delta_{\mathrm{max}}=\pi/(50I)$ and a typical frame
size of $L_d=3L_{d1}=1536$ bits or $L_d I$ samples, the phase change due
the frequency offset over the entire frame is
%*******************************************************************************
\begin{align}
\label{Eq:Pap19_Eq14}
\Delta\theta & = \frac{\pi}{50I}
                 \times
                  L_d I    \nonumber  \\
             & = 96.51 \quad \mbox{radians}
\end{align}
%*******************************************************************************
which is significant. Therefore it is necessary to estimate and cancel the
frequency offset. In what follows, we describe the optimum (maximum likelihood
(ML)) joint frequency offset and frame detection procedure. Let
%*******************************************************************************
\begin{equation}
\label{Eq:Pap19_Eq14_0}
S_{k,\, t} = \left
             \{
             \begin{array}{ll}
                 S_k &  \mbox{if $S_k$ is training bit}\\
                 0   &  \mbox{otherwise.}
             \end{array}
             \right.
\end{equation}
%*******************************************************************************
In other words, $S_{k,\, t}$ contains only training bits, with data and
parity bits set to zero. Define
%*******************************************************************************
\begin{align}
\label{Eq:Pap19_Eq14_0_1}
 s_{t,\, I}(mT_s) & = \sum_{k=0}^{L_d-1}
                       S_{k,\, t}
                       p(mT_s-kT)                  \nonumber  \\
 s_{t,\, Q}(mT_s) & = \sum_{k=0}^{L_d-1}
                       S_{k,\, t}
                       \hat{p}_\mathrm{m}(mT_s-kT).
\end{align}
%*******************************************************************************
Note that $s_{t,\, I}(mT_s)$ and $s_{t,\, Q}(mT_s)$ can be precomputed and
stored at the receiver. The joint detection is given by
%*******************************************************************************
\begin{align}
\label{Eq:Pap19_Eq14_1}
\max_{\hat{\omega}_i,\, m}
 Z\left(\hat{\omega}_i,\, m\right)
& =
\left|
\left(
 2r(mT_s)\cos
         \left(
         \hat{\omega}_i m
         \right)
\right)\star s_{t,\, I}(-mT_s)
\right.                          \nonumber  \\
&
\quad -
\left.
\mathrm{j}\,
\left(
 2r(mT_s)\sin
         \left(
         \hat{\omega}_i m
         \right)
\right)\star s_{t,\, Q}(-mT_s)
\right|
\end{align}
%*******************************************************************************
where
%*******************************************************************************
\begin{equation}
\label{Eq:Pap19_Eq14_2}
\hat{\omega}_i = \frac{2\pi k_i}{I} + \hat{\delta}_i.
\end{equation}
%*******************************************************************************
Note that (\ref{Eq:Pap19_Eq14_1}) corresponds to demodulation followed by
matched filtering by the baseband training signals given in
(\ref{Eq:Pap19_Eq14_0_1}). The signals $s_{t,\, I}(-mT_s)$ and
$s_{t,\, Q}(-mT_s)$ in (\ref{Eq:Pap19_Eq14_1}) are referred to as
``baseband training matched filters''.
%*******************************************************************************
\begin{figure}[tbhp]
\begin{center}
%\centering
\input{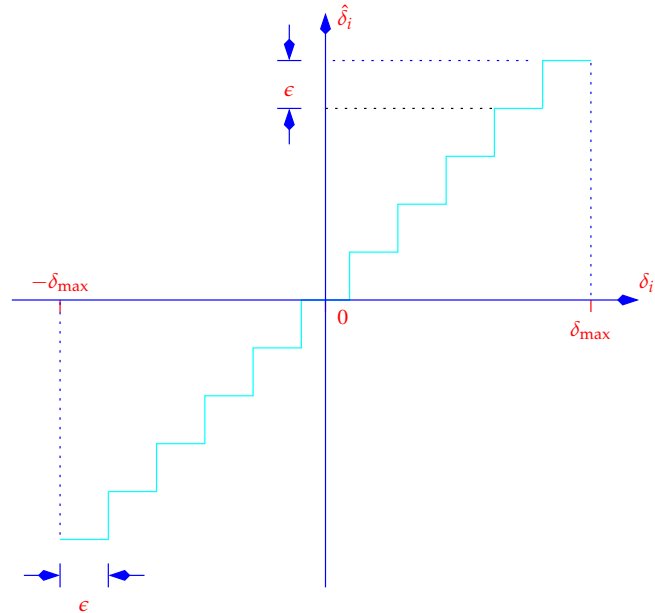}
\caption{Mid-step quantizer for frequency offset estimation.}
\label{Fig:FOFF_Est}
\end{center}
\end{figure}
%*******************************************************************************
The frequency offset is estimated using a uniform mid-step quantizer 
(a uniform mid-rise quantizer can also be used) \cite{Haykin83,Vasu_AC_PS}.
The interval $\left[-\delta_{\mathrm{max}},\, \delta_{\mathrm{max}}\right]$
is divided into $B$ intervals, where $B$ is an odd integer, as shown in
Figure~\ref{Fig:FOFF_Est}, for $B=11$. The step-size is
%*******************************************************************************
\begin{equation}
\label{Eq:Pap19_Eq15}
\epsilon = \frac{2\delta_{\mathrm{max}}}{B} \quad \mbox{radians}.
\end{equation}
%*******************************************************************************
Clearly from Figure~\ref{Fig:FOFF_Est}, the maximum error in the frequency
offset estimate is $\pm\epsilon/2$. We restrict the phase change due to error in
the frequency offset estimate (residual frequency offset) over the whole
frame to $0.1\pi$. Hence we obtain
%*******************************************************************************
\begin{align}
\label{Eq:Pap19_Eq16}
\frac{\epsilon}{2} L_d I
& = 0.1\pi \quad \mbox{radians}   \nonumber  \\
\Rightarrow
\frac{\delta_{\mathrm{max}}}{B} L_d I
& = 0.1\pi                        \nonumber  \\
\Rightarrow
\frac{\pi}{50IB} L_d I
& = 0.1\pi                        \nonumber  \\
\Rightarrow
 B
& = \frac{L_d}{5}.
\end{align}
%*******************************************************************************
For a typical value of $L_d=1536$ bits, we get $B=307.2\equiv 309$
(next higher odd integer) intervals. Thus with $I=16$, the complexity of
(\ref{Eq:Pap19_Eq14_1}) is of the order of $\mathscr{O}(L_dIB)=7,593,984$
operations (additions, subtractions and multiplications) for each subcarrier,
which is quite large. Therefore, we next describe the two-step approach for
frequency offset estimation, which is much more efficient. In the first step,
namely coarse, we constrain the phase change due to residual frequency
offset over the entire frame to
%*******************************************************************************
\begin{align}
\label{Eq:Pap19_Eq17}
\frac{\epsilon_c}{2} L_d I
& = \pi \quad \mbox{radians}       \nonumber  \\
\Rightarrow
\frac{\delta_{\mathrm{max}}}{B_c} L_d I
& = \pi                            \nonumber  \\
\Rightarrow
\frac{\pi}{50IB_c} L_d I
& = \pi                            \nonumber  \\
\Rightarrow
 B_c
& = \frac{L_d}{50}
\end{align}
%*******************************************************************************
where
%*******************************************************************************
\begin{equation}
\label{Eq:Pap19_Eq17_1}
\epsilon_c = \frac{2\delta_{\mathrm{max}}}{B_c} \quad \mbox{radians}.
\end{equation}
%*******************************************************************************
Using $L_d=1536$, we get $B_c=30.72\equiv 31$ (odd integer).
Clearly, the savings in computation in (\ref{Eq:Pap19_Eq17}) over
(\ref{Eq:Pap19_Eq16}) is by a factor of
%*******************************************************************************
\begin{equation}
\label{Eq:Pap19_Eq18}
\frac{B}{B_c} = 10.
\end{equation}
%*******************************************************************************
%*******************************************************************************
\begin{figure}[tbhp]
\begin{center}
%\centering
\input{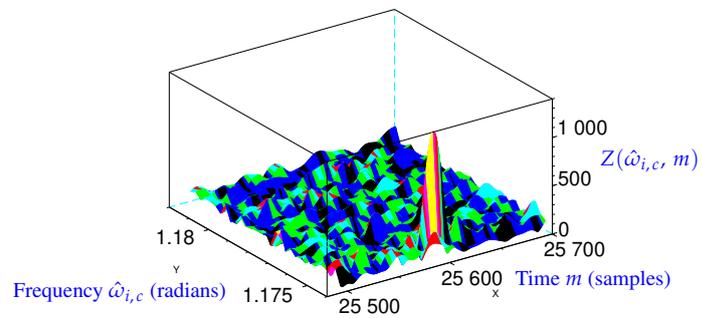}
\caption{Coarse frequency offset and frame detection.}
\label{Fig:DMT63e_Coarse_FOFF}
\end{center}
\end{figure}
%*******************************************************************************
The result of coarse frequency offset estimate obtained from
(\ref{Eq:Pap19_Eq14_1}), using $B_c$ intervals, is shown in
Figure~\ref{Fig:DMT63e_Coarse_FOFF}, at an average SNR per bit of zero dB,
and random frequency offset in the range given in (\ref{Eq:Pap19_Eq7}).
The interpolation factor is $I=16$, the index of the second subcarrier is
$k_2=3$ and the roll-off factor of the RRC spectrum is $\rho=0.161$.
Hence, the subcarrier frequency is $2\pi k_2/I=1.1781$ radians and
from (\ref{Eq:Pap19_Eq3_1}) we obtain the
signal bandwidth as 0.228 radians on one (upper) side the subcarrier
frequency.
%*******************************************************************************
\begin{figure}[tbhp]
\begin{center}
%\centering
\input{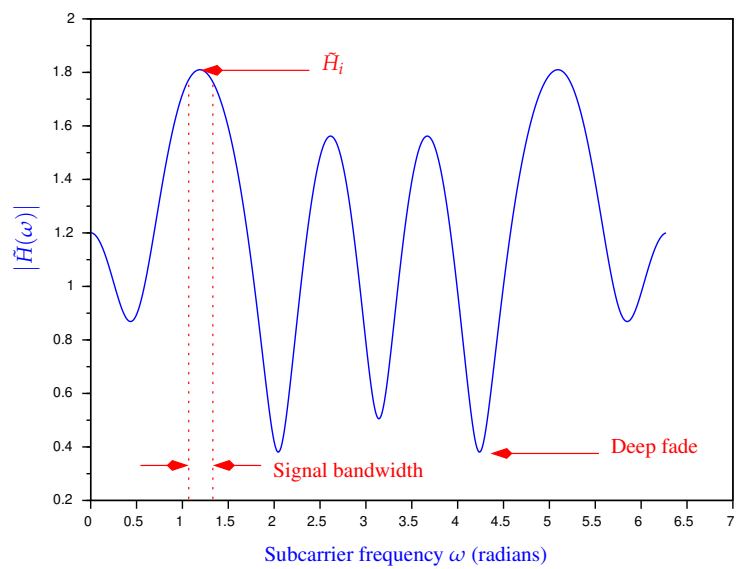}
\caption{Channel magnitude response from 0 to $2\pi$ radians.}
\label{Fig:Chan_Mag_FOFF}
\end{center}
\end{figure}
%*******************************************************************************
The channel magnitude response used to obtain
Figure~\ref{Fig:DMT63e_Coarse_FOFF},
is shown in Figure~\ref{Fig:Chan_Mag_FOFF}, with $\omega=2\pi l/512$, for
$0\le l < 512$ and channel length $L_h=8$ coefficients
(see also (\ref{Eq:Pap19_Eq11})). The signal bandwidth is also
shown in Figure~\ref{Fig:Chan_Mag_FOFF}. Note that for coarse frequency offset
estimation, $\epsilon$ in Figure~\ref{Fig:FOFF_Est} must be replaced by
$\epsilon_c$ and $\hat{\delta}_i$ must be replaced by
$\hat{\delta}_{i,\, c}$. Consequently, $\hat{\omega}_i$ in
(\ref{Eq:Pap19_Eq14_1}) must be replaced by $\hat{\omega}_{i,\, c}$, where
%*******************************************************************************
\begin{equation}
\label{Eq:Pap19_Eq18_1}
\hat{\omega}_{i,\, c} = \frac{2\pi k_i}{I} + \hat{\delta}_{i,\, c}.
\end{equation}
%*******************************************************************************

Let $\hat{\delta}_{i,\, c}$ denote the
coarse estimate of the frequency offset obtained from (\ref{Eq:Pap19_Eq14_1}),
using $B_c$ intervals of size $\epsilon_c$. Next, fine frequency offset
estimate needs to be obtained. This is done by considering the residual
frequency offset to be in the range
$[\hat{\delta}_{i,\, c}-\epsilon_c,\, \hat{\delta}_{i,\, c}+\epsilon_c]$
and dividing this range into $B_f=21$ intervals. If the actual frequency
offset $\delta_i$ lies outside this range, it is said to be an
\textit{outlier}.
The size of each interval during fine frequency offset estimation is
%*******************************************************************************
\begin{align}
\label{Eq:Pap19_Eq19}
\epsilon_f & = \frac{2\epsilon_c}{B_f}     \nonumber  \\
           & = \frac{2\times 2\delta_{\mathrm{max}}}{B_c B_f}
               \quad \mbox{radians.}
\end{align}
%*******************************************************************************
Assuming an error in the  fine frequency offset estimate to be
$\pm \epsilon_f/2$, the phase change due to the residual frequency offset
estimate over the entire frame is
%*******************************************************************************
\begin{align}
\label{Eq:Pap19_Eq20}
\frac{\epsilon_f}{2} L_d I
& =
\frac{2\delta_{\mathrm{max}}}{B_c B_f} L_d I          \nonumber  \\
& =
\frac{2\pi}{50IB_cB_f} L_d I                          \nonumber  \\
& = 0.09\pi                                           \nonumber  \\
& < 0.1\pi
\end{align}
%*******************************************************************************
for $L_d=1536$, $B_c=31$ and $B_f=21$.
%*******************************************************************************
\begin{figure}[tbhp]
\begin{center}
%\centering
\input{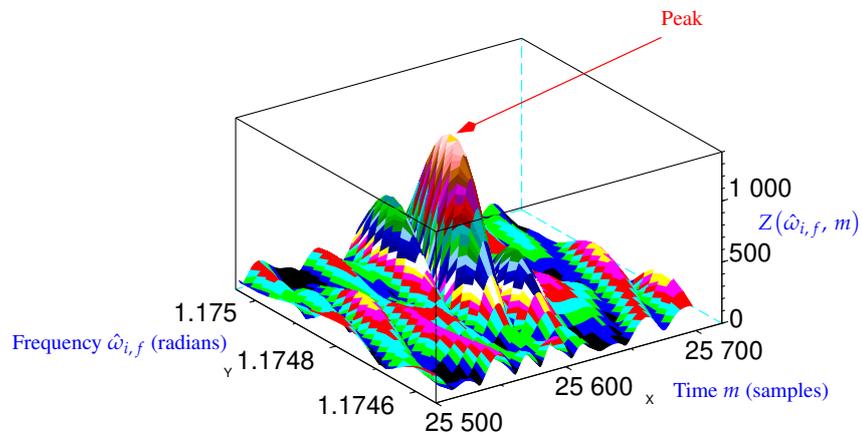}
\caption{Fine frequency offset and frame detection.}
\label{Fig:DMT63e_Fine_FOFF}
\end{center}
\end{figure}
%*******************************************************************************
The result of fine frequency offset estimation and frame detection
using (\ref{Eq:Pap19_Eq14_1}) is shown in Figure~\ref{Fig:DMT63e_Fine_FOFF},
at an average SNR per bit equal to zero dB. Observe that for fine frequency
offset estimation, $\delta_{\mathrm{max}}$ in Figure~\ref{Fig:FOFF_Est}
must be replaced by $\hat{\delta}_{i,\, c}+\epsilon_c$,
$-\delta_{\mathrm{max}}$ must be replaced by
$\hat{\delta}_{i,\, c}-\epsilon_c$, $\epsilon$ must be replaced by
$\epsilon_f$ as given in (\ref{Eq:Pap19_Eq19}) and $\hat{\delta}_i$
must be replaced by $\hat{\delta}_{i,\, f}$. Similarly,
$\hat{\omega}_i$ in (\ref{Eq:Pap19_Eq14_1}) must be replaced by
$\hat{\omega}_{i,\, f}$ where
%*******************************************************************************
\begin{equation}
\label{Eq:Pap19_Eq20_0}
\hat{\omega}_{i,\, f} = \hat{\omega}_{i,\, c} + \hat{\delta}_{i,\, f}
\end{equation}
%*******************************************************************************
where $\hat{\omega}_{i,\, c}$ is given in (\ref{Eq:Pap19_Eq18_1}).

We conclude this subsection with the following observations:
%%%%%%%%%%%%%%%%%%%%%%%%%%%%%%%%%%%%%%%%%%%%%%%%%%%%%%%%%%%%%%%%%%%%%%%%%%%%%%%%
\begin{enumerate}
    \item The joint frequency offset and frame detection procedure in
          (\ref{Eq:Pap19_Eq14_1}) has a lot of scope for parallel processing.
    \item Observe that $\hat{\omega}_i$ in (\ref{Eq:Pap19_Eq14_1}) must be
          replaced by $\hat{\omega}_{i,\, c}$ for coarse frequency offset
          estimation and by $\hat{\omega}_{i,\, f}$ for fine frequency offset
          estimation.
    \item The frame is ``detected'' during fine frequency offset estimation
          by computing the peak-to-average (PTA)
          value of $Z^2\left(\hat{\omega}_{i,\, f},\, m\right)$ over all
          $\hat{\omega}_{i,\, f}$ and $m$ in (\ref{Eq:Pap19_Eq14_1}).
          The frame is declared as ``detected'' if $\mbox{PTA} > 100$ for
          at least one subcarrier.
          Conversely, the frame transmitted over a subcarrier in deep
          fade, as illustrated in Figure~\ref{Fig:Chan_Mag_FOFF}, may not
          be detected, if $\mbox{PTA}<100$. If all the subcarriers are
          simultaneously in deep fade, then the frame may not be detected.
          This is referred to as an ``erasure'' \cite{KV_SSID2020}.
    \item The computational complexity per subcarrier for the two-stage
          approach is
%*******************************************************************************
\begin{equation}
\label{Eq:Pap19_Eq20_1}
\mathscr{O}\left(L_d I(B_c+B_f)\right)
\end{equation}
%*******************************************************************************
          whereas for the single-stage approach it is $\mathscr{O}(L_d I B)$.
          Hence, the savings in computation is by a factor of
%*******************************************************************************
\begin{equation}
\label{Eq:Pap19_Eq21}
\frac{B}{B_c+B_f} = 5.9
\end{equation}
%*******************************************************************************
          for $B=309$, $B_c=31$ and $B_f=21$.
    \item Once the frame is ``detected'', it is straightforward to compute
          the sampling instants $t=kT$ at the output of the two matched
          filters (see Figure~\ref{Fig:OFDM_OQAM_Sys_Model1}),
          in order to obtain (\ref{Eq:Pap19_Eq10}), and accounting for the
          training bits. Since the ``peak''
          (see Figure~\ref{Fig:DMT63e_Fine_FOFF}) is obtained at the end
          of the frame (recall that the first and last bits of the frame
          correspond to training, as shown in
          Figure~\ref{Fig:OFDM_OQAM_Frame1}), it coincides with the last
          training bit. Therefore, the remaining bits in the frame can
          be obtained by starting from the ``peak'', going backwards
          in time, and downsampling the matched filter outputs by a factor
          of $I$.
    \item The root mean squared (RMS) coarse frequency offset estimation error is
          given by
%*******************************************************************************
\begin{equation}
\label{Eq:Pap19_Eq20_0_1}
\omega_{\mathrm{e,\, rms},\, c} = \left(\,\overline{
                                  \left(
                                  \omega_i -
                                  \hat{\omega}_{i,\, c}
                                  \right)^2
                                  }
                                  \,
                                  \right)^{1/2}
\end{equation}
%*******************************************************************************
          where $\overline{\left(\cdot\right)}$ denotes time average
          over all frames
          and subcarriers, $\omega_i$ is given in (\ref{Eq:Pap19_Eq6}) and
          $\hat{\omega}_{i,\, c}$ is given in (\ref{Eq:Pap19_Eq18_1}).
          Similarly, the RMS fine frequency offset estimation error is
%*******************************************************************************
\begin{equation}
\label{Eq:Pap19_Eq20_0_2}
\omega_{\mathrm{e,\, rms},\, f} =
                                 \left(\,\overline{
                                 \left(
                                 \omega_i -
                                 \hat{\omega}_{i,\, f}
                                 \right)^2
                                 }
                                 \,
                                 \right)^{1/2}
\end{equation}
%*******************************************************************************
          where, the time average is taken over all \textit{detected} frames
          and subcarriers and $\hat{\omega}_{i,\, f}$ is given in
          (\ref{Eq:Pap19_Eq20_0}).
\end{enumerate}
%%%%%%%%%%%%%%%%%%%%%%%%%%%%%%%%%%%%%%%%%%%%%%%%%%%%%%%%%%%%%%%%%%%%%%%%%%%%%%%%
In the next subsection, we discuss channel and noise variance estimation.
%%%%%%%%%%%%%%%%%%%%%%%%%%%%%%%%%%%%%%%%%%%%%%%%%%%%%%%%%%%%%%%%%%%%%%%%%%%%%%%%
\subsection{Channel and Noise Variance Estimation}
\label{SSec:Chan_Noise_Est}
%%%%%%%%%%%%%%%%%%%%%%%%%%%%%%%%%%%%%%%%%%%%%%%%%%%%%%%%%%%%%%%%%%%%%%%%%%%%%%%%
From Figure~\ref{Fig:Chan_Mag_FOFF}, it is clear that the signal bandwidth
is narrow enough for the channel to be considered ideal (channel magnitude
response is flat and phase response is linear). Therefore, it is only
necessary to estimate $\tilde{H}_i$ in (\ref{Eq:Pap19_Eq10}) and
(\ref{Eq:Pap19_Eq11}). Observe that the term inside the magnitude in
(\ref{Eq:Pap19_Eq14_1}) corresponding to the peak
(see Figure~\ref{Fig:DMT63e_Fine_FOFF}) during
fine frequency offset estimation is equal to (see also (\ref{Eq:Pap19_Eq10}))
%*******************************************************************************
\begin{equation}
\label{Eq:Pap19_Eq22}
\tilde{Z}_{\mathrm{max}}\left(\hat{\omega}_{i,\, f},\, m\right)
=  L_{d1}
  \tilde{H}_i (1+\mathrm{j}) + \tilde{z}_{1,\, i,\, m}
\end{equation}
%*******************************************************************************
where it is assumed that there are $L_{d1}$ training bits, the filters
$p(mT_s)$ and $\hat{p}_{\mathrm{m}}(mT_s)$ have unit energy and
$\tilde{z}_{1,\, i,\, m}$ denotes complex valued noise. It can be shown that
the variance per dimension of $\tilde{z}_{1,\, i,\, m}$ is
(see also (\ref{Eq:Pap19_Eq12}))
%*******************************************************************************
\begin{equation}
\label{Eq:Pap19_Eq22_1}
\frac{1}{2}
 E
\left[
\left|
\tilde{z}_{1,\, i,\, m}
\right|^2
\right] =  2L_{d1}
          \sigma^2_w.
\end{equation}
%*******************************************************************************
Observe that since (\ref{Eq:Pap19_Eq22}) is the peak output of the
training matched filter, the SNR is maximized \cite{Vasu_Book10}.
The statement of the maximum likelihood (ML) channel estimation problem is as
follows \cite{6663392,Vasudevan2015,Vasu_Adv_Tele_2017,Vasu_intech:2019}:
find $\hat{\tilde{H}}_i$ such that
%*******************************************************************************
\begin{equation}
\label{Eq:Pap19_Eq23}
\left|
\tilde{Z}_{\mathrm{max}}\left(\hat{\omega}_{i,\, f},\, m\right)
-  L_{d1}
\hat{\tilde{H}}_i (1+\mathrm{j})
\right|^2
\end{equation}
%*******************************************************************************
is minimized. Differentiating (\ref{Eq:Pap19_Eq23}) with
respect to $\hat{\tilde{H}}_i^*$ \cite{Vasu_Book10} and setting the result
to zero, we obtain
%*******************************************************************************
\begin{equation}
\label{Eq:Pap19_Eq24}
\hat{\tilde{H}}_i =
\frac{
\tilde{Z}_{\mathrm{max}}\left(\hat{\omega}_{i,\, f},\, m\right)}
{(1+\mathrm{j})L_{d1}}.
\end{equation}
%*******************************************************************************
The RMS channel estimation error is given by
%*******************************************************************************
\begin{equation}
\label{Eq:Pap19_Eq24_1}
 H_{\mathrm{e,\, rms}} = \left(
                         \,
                         \overline{
                         \left(
                         \left|
                         \tilde{H}_i
                         \right| -
                         \left|
                         \hat{\tilde{H}}_i
                         \right|
                         \right)^2
                         }
                         \,
                         \right)^{1/2}
\end{equation}
%*******************************************************************************
where the time average is taken over all detected frames and subcarriers,
$\hat{\tilde{H}}_i$ is given in (\ref{Eq:Pap19_Eq24}) and $\tilde{H}_i$
is given in (\ref{Eq:Pap19_Eq11}). Observe that the difference of the
absolute values is taken in (\ref{Eq:Pap19_Eq24_1}), since
$\hat{\tilde{H}}_i$ may contain an unknown carrier phase.

The next step is to estimate the two-dimensional noise variance at the
matched filter output, since it is required for turbo decoding
\cite{Vasu_Book10}. Let $k_t$ denote the time instant corresponding training
bit. Then the estimate of the one-dimensional noise variance for the
$i^{th}$ subcarrier is given by
%*******************************************************************************
\begin{equation}
\label{Eq:Pap19_Eq25}
\hat{\sigma}^2_{w,\, i} =
\frac{1}{2L_{d1}}
\sum_{\substack{k=0\\k\in k_t}}^{L_d-1}
\left|
\tilde{x}_{i,\, k} -
\hat{\tilde{H}}_i
(1+\mathrm{j}) S_{k,\, t}
\right|^2
\end{equation}
%*******************************************************************************
where $\hat{\tilde{H}}_i$ is given in (\ref{Eq:Pap19_Eq24}) and $S_{k,\, t}$
is given in (\ref{Eq:Pap19_Eq14_0}). Note that the estimate of the
one-dimensional noise variance at the output of the matched filters is
$2\hat{\sigma}^2_{w,\, i}$ (see also (\ref{Eq:Pap19_Eq12})).
The RMS error in the estimation of the two-dimensional noise variance at the
matched filter outputs is
%*******************************************************************************
\begin{equation}
\label{Eq:Pap19_Eq25_1}
\sigma^2_{\mathrm{e,\, rms,\, MFop}} = \left(
                                       \,
                                       \overline{
                                       \left(
                                        4
                                       \sigma^2_w -
                                        4
                                       \hat{\sigma}^2_{w,\, i}
                                       \right)^2
                                       }
                                       \,
                                       \right)^{1/2}
\end{equation}
%*******************************************************************************
where the time average is taken over all detected frames and subcarriers.
In the next subsection, we briefly
discuss turbo decoding using the BCJR algorithm \cite{BCJR74,Vasu_Book10}.
%%%%%%%%%%%%%%%%%%%%%%%%%%%%%%%%%%%%%%%%%%%%%%%%%%%%%%%%%%%%%%%%%%%%%%%%%%%%%%%%
\subsection{Turbo Decoding}
\label{SSec:Turbo_Decode}
%%%%%%%%%%%%%%%%%%%%%%%%%%%%%%%%%%%%%%%%%%%%%%%%%%%%%%%%%%%%%%%%%%%%%%%%%%%%%%%%
Turbo decoding using the BCJR algorithm has been discussed extensively in
\cite{Vasu_Book10}, and hence will not be repeated here, except for
specific details. We begin by observing that the data and parity bits
are distributed across random time slots in the frame, since the training
bits are also at random time slots, other than the first and last slot.
Hence, for the purpose of turbo decoding, the data and parity bits have
to be re-arranged to consecutive time slots, denoted by $k_r$, for
$0\le k_r\le L_{d1}-1$. We assume that this has been done. Let
$\tilde{x}_{i,\, k_r,\, b1}$
denote the data sample, $\tilde{x}_{i,\, k_r,\, c1}$ denote the parity
sample  from encoder 1 and $\tilde{x}_{i,\, k_r,\, c2}$ the parity sample
from encoder 2, all extracted and re-arranged from $\tilde{x}_{i,\, k}$
in (\ref{Eq:Pap19_Eq10}). Since a rate-$1/2$ turbo code is used,
$\tilde{x}_{i,\, k_r,\, c1}$ is defined only for even values of $k_r$,
whereas $\tilde{x}_{i,\, k_r,\, c2}$ is defined only for odd values
of $k_r$. Following the
notation in \cite{Vasu_Book10}, for a trellis transition at time $k_r$,
from encoder state $e$ to $f$ at decoder 1
%*******************************************************************************
\begin{equation}
\label{Eq:Pap19_Eq26}
\gamma_{1,\,\mathrm{sys},\, k_r,\, e,\, f}  = \exp
                                              \left[-
                                               d^2_{k_r,\, b1,\, e,\, f}
                                              \right]
\end{equation}
%*******************************************************************************
where
%*******************************************************************************
\begin{equation}
\label{Eq:Pap19_Eq27}
 d^2_{k_r,\, b1,\, e,\, f}    = \sum_{i=1}^{N_{\mathrm{sc,\, d}}}
                               \frac{
                               \left|
                               \tilde{x}_{i,\, k_r,\, b1} -
                               \hat{\tilde{H}}_i
                                S_{b,\, e,\, f}
                               \right|^2}
                               {2\hat{\sigma}^2_{w,\, i}}           
\end{equation}
%*******************************************************************************
where $N_{\mathrm{sc,\, d}}\le N_{\mathrm{sc}}$ is the number of detected
subcarriers, $S_{b,\, e,\, f}\in\pm 1$ is the systematic (data) bit
corresponding to
the transition from encoder state $e$ to $f$, $\hat{\tilde{H}}_i$ is given in
(\ref{Eq:Pap19_Eq24}) and
$\hat{\sigma}_{w,\, i}^2$ is given in (\ref{Eq:Pap19_Eq25}). Similarly
%*******************************************************************************
\begin{equation}
\label{Eq:Pap19_Eq28}
\gamma_{1,\,\mathrm{par},\, k_r,\, e,\, f} =  \left
                                              \{
                                              \begin{array}{ll}
                                              \exp
                                              \left[-
                                               d^2_{k_r,\, c1,\, e,\, f}
                                              \right] & \mbox{for even $k_r$}\\
                                               1      & \mbox{otherwise}
                                              \end{array}
                                              \right.
\end{equation}
%*******************************************************************************
where
%*******************************************************************************
\begin{equation}
\label{Eq:Pap19_Eq29}
 d^2_{k_r,\, c1,\, e,\, f} =   \sum_{i=1}^{N_{\mathrm{sc,\, d}}}
                               \frac{
                               \left|
                               \tilde{x}_{i,\, k_r,\, c1} -
                               \hat{\tilde{H}}_i
                                S_{c,\, e,\, f}
                               \right|^2}
                               {2\hat{\sigma}^2_{w,\, i}}        
\end{equation}
%*******************************************************************************
where $S_{c,\, e,\, f}\in\pm 1$ is the parity bit corresponding to the
transition from encoder state $e$ to $f$. The equations for decoder 2 are
similar, excepting that $\tilde{x}_{i,\, k_r,\, b2}$ and
$\tilde{x}_{i,\, k_r,\, c2}$ have to be used. Note that
$\tilde{x}_{i,\, k_r,\, b2}$ is the interleaved version of
$\tilde{x}_{i,\, k_r,\, b1}$. Moreover
$\gamma_{2,\,\mathrm{par},\, k_r,\, e,\, f}$ has to be set to unity for
even $k_r$, unlike $\gamma_{1,\,\mathrm{par},\, k_r,\, e,\, f}$ in
(\ref{Eq:Pap19_Eq28}). In the next section we discuss computer simulation
results.
%%%%%%%%%%%%%%%%%%%%%%%%%%%%%%%%%%%%%%%%%%%%%%%%%%%%%%%%%%%%%%%%%%%%%%%%%%%%%%%%
\section{Results}
\label{Sec:Results}
%%%%%%%%%%%%%%%%%%%%%%%%%%%%%%%%%%%%%%%%%%%%%%%%%%%%%%%%%%%%%%%%%%%%%%%%%%%%%%%%
%*******************************************************************************
\begin{table}[tbhp]
\begin{center}
%\centering
\caption{Simulation parameters.}
\input{pap19_sim_param.pstex_t}
\label{Tbl:Pap19_Sim_Param}
\end{center}
\end{table}
%*******************************************************************************
The simulation parameters are shown in Table~\ref{Tbl:Pap19_Sim_Param}.
%*******************************************************************************
\begin{figure}[tbhp]
\begin{center}
%\centering
\input{dmt63e_plt_nsc1.pstex_t}
\caption{BER results for $N_{\mathrm{sc}}=1$.}
\label{Fig:DMT63e_Plt_Nsc1}
\end{center}
\end{figure}
%*******************************************************************************
The BER results for $N_{\mathrm{sc}}=1$ is shown in
Figure~\ref{Fig:DMT63e_Plt_Nsc1}. Here ``\texttt{C\_BER}'' denotes the
turbo coded BER from simulations, ``\texttt{T\_BER}'' is the theoretical BER
from \cite{Vasudevan23}, ``\texttt{U\_BER}'' is the simulated uncoded BER and
``\texttt{ERASE}'' is the probability of erasure \cite{KV_SSID2020}. Observe
the close match between \texttt{C\_BER} and \texttt{T\_BER}.
%*******************************************************************************
\begin{figure}[tbhp]
\begin{center}
%\centering
\input{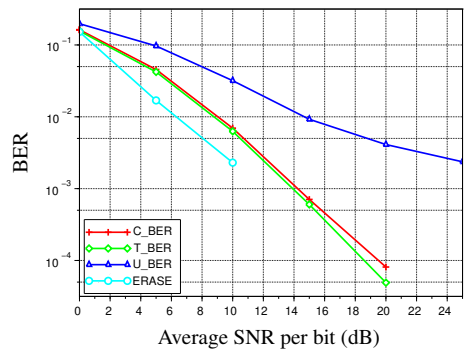}
\caption{BER results for $N_{\mathrm{sc}}=2$.}
\label{Fig:DMT63e_Plt_Nsc2}
\end{center}
\end{figure}
%*******************************************************************************
%*******************************************************************************
\begin{figure}[tbhp]
\begin{center}
%\centering
\input{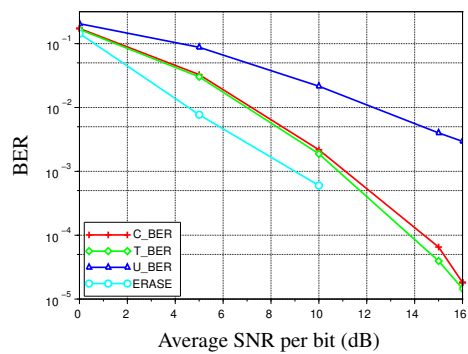}
\caption{BER results for $N_{\mathrm{sc}}=3$.}
\label{Fig:DMT63e_Plt_Nsc3}
\end{center}
\end{figure}
%*******************************************************************************
%*******************************************************************************
\begin{figure}[tbhp]
\begin{center}
%\centering
\input{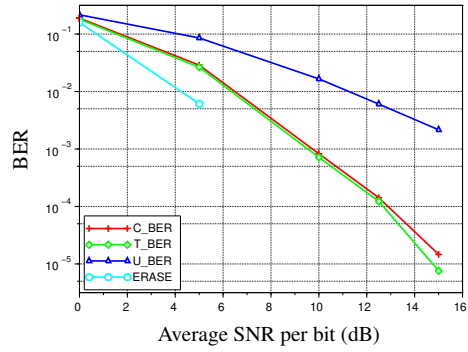}
\caption{BER results for $N_{\mathrm{sc}}=4$.}
\label{Fig:DMT63e_Plt_Nsc4}
\end{center}
\end{figure}
%*******************************************************************************
The simulation results for $N_{\mathrm{sc}}=2,\, 3,\, 4$ are presented in
Figures~\ref{Fig:DMT63e_Plt_Nsc2}-\ref{Fig:DMT63e_Plt_Nsc4}. The following
observations can be made from 
Figures~\ref{Fig:DMT63e_Plt_Nsc1}-\ref{Fig:DMT63e_Plt_Nsc4}:
%*******************************************************************************
\begin{enumerate}
    \item At a BER of $10^{-4}$, there is about 10 dB improvement between
          $N_{\mathrm{sc}}=1,\, 2$, about 6 dB improvement between
          $N_{\mathrm{sc}}=2,\, 3$ and only about 1 dB improvement between
          $N_{\mathrm{sc}}=3,\, 4$. This shows that increasing
          $N_{\mathrm{sc}}$ has diminishing returns.
    \item The BER performance may be limited by ISI, since the channel
          is not distortionless. The BER performance may be improved by
          increasing the interpolation factor $I$, for a fixed channel
          length $L_h$. This would reduce the signal bandwidth in
          Figure~\ref{Fig:Chan_Mag_FOFF}, which would better approximate
          the distortionless channel assumption. However, increasing
          $I$ results in increased simulation complexity
          (see (\ref{Eq:Pap19_Eq20_1})).
    \item Increasing the subcarrier diversity $N_{\mathrm{sc}}$ results in
          increased transmission bandwidth, since it is equivalent to
          frequency division multiplexing. Perhaps antenna diversity is
          an alternate solution. Therefore, the BER performance of the
          proposed OFDM-OQAM system can be studied for MIMO systems.
    \item The BER performance of turbo coded OFDM-OQAM is far superior to
          that of uncoded.
\end{enumerate}
%*******************************************************************************
%*******************************************************************************
\begin{table}[tbhp]
\begin{center}
%\centering
\caption{RMS error at an average SNR per bit of 5 dB. Values in red colour
         correspond to ideal channel.}
\input{rms_err.pstex_t}
\label{Tbl:RMS_Err}
\end{center}
\end{table}
%*******************************************************************************
The RMS estimation error in varous parameters is given in
Table~\ref{Tbl:RMS_Err}, at an average SNR per bit of 5 dB. The numbers
in red colour correspond to ideal channel, that is
%*******************************************************************************
\begin{equation}
\label{Eq:Pap19_Eq30}
h(nT_s) = \delta_K(nT_s).       
\end{equation}
%*******************************************************************************
The numbers in blue are for fading channel as in (\ref{Eq:Pap19_Eq9}).
We can make the following observations from Table~\ref{Tbl:RMS_Err}:
%*******************************************************************************
\begin{enumerate}
 \item The RMS estimation errors for the fading channel is much
       higher than the ideal channel. This can be attributed to the
       ISI introduced by the fading channel.
 \item The \texttt{C\_BER} of the ideal channel is also much lower than
       that of the fading channels
       (Figures~\ref{Fig:DMT63e_Plt_Nsc1}-\ref{Fig:DMT63e_Plt_Nsc4}),
       at an SNR per bit of 5 dB.
       In fact, for $N_{\mathrm{sc}}=1$, there were zero errors out of
       $512\times 10^4$ bits (less than one error out of
       $512\times 10^4$ bits). Similarly, for $N_{\mathrm{sc}}=2,\, 3,\, 4$
       there were less than 10 errors out of $512\times 10^4$ bits, hence
       the \texttt{C\_BER} is given as an approximation. This again
       indicates that the \texttt{C\_BER} for fading channels is limited by
       ISI.
\end{enumerate}
%*******************************************************************************
It was also observed from computer simulations that the RMS estimation
errors for fading channels, as given in Table~\ref{Tbl:RMS_Err},
is quite insensitive to wide variations in the SNR ($0-25$ dB), which
can be attributed to ISI.
%*******************************************************************************
\section{Conclusions}
\label{Sec:Conclude}
A new approach to OFDM-OQAM using the concept of single sideband modulation
(SSB) has been discussed. The pulse corresponding to the square-root of the
raised cosine (SRRC) spectrum and its modified Hilbert transform has been used
to obtain bandwidth reduction by a factor of two. A simple matched filter is
used at the receiver. The bit-error-rate (BER)
performance of the proposed OFDM-OQAM system has been studied under various
impairments such as carrier frequency offset (CFO), frequency selective
Rayleigh fading channel and additive white Gaussian noise (AWGN).
Discrete-time algorithms for frame detection, two-step CFO estimation,
channel and noise variance estimation have been proposed.
A single transmit and receive antenna is assumed. Similar work has not been done
earlier. Future work could involve MIMO systems.

\conflictsofinterest{The author declares no conflicts of interest.} 

\begin{adjustwidth}{-\extralength}{0cm}

\reftitle{References}


\bibliography{mybib,mybib1,mybib2,mybib3,mybib4,mybib5}

\PublishersNote{}
%\isPreprints{}{% This command is only used for ``preprints''.
\end{adjustwidth}
\end{document}